\documentclass{aa}
\usepackage{graphicx}

\begin{document}

\title{Confirmation of persistent radio jets in the microquasar \object{LS~5039}}

\author{J.~M. Paredes\inst{1}
\and M. Rib\'o\inst{1}
\and E. Ros\inst{2}
\and J. Mart\'{\i}\inst{3}
\and M. Massi\inst{2}
}

\offprints{J.~M. Paredes, \\ \email{josep@am.ub.es}}

\institute{Departament d'Astronomia i Meteorologia, Universitat de Barcelona, Av. Diagonal 647, 08028 Barcelona, Spain
\and Max-Planck-Institut f\"ur Radioastronomie, Auf dem H\"ugel 69, 53121 Bonn, Germany
\and Departamento de F\'{\i}sica, Escuela Polit\'ecnica Superior, Universidad de Ja\'en, Virgen de la Cabeza 2, 23071 Ja\'en, Spain
}

\date{Received / Accepted}


\abstract{
We present here new observations conducted with the EVN and MERLIN of the
persistent microquasar \object{LS~5039} discovered by Paredes et~al.
(\cite{paredes00}) with the VLBA. The new observations confirm the presence of
an asymmetric two-sided jet reaching up to $\sim1000$~AU on the longest jet
arm. The results suggest a bending of the jets with increasing distance from
the core and/or precession. The origin and location of the high-energy
gamma-ray emission associated with the system is discussed and an estimate of
the magnetic field at the base of the jet given. Our results suggest a well
collimated radio jet. We also comment on new observing strategies to be used
with satellites and forthcoming detectors, since this persistent source
appears to be a rather good laboratory to explore the accretion/ejection
processes taking place near compact objects.
}

\maketitle

\section{Introduction} \label{sec:introduction}

Microquasars are radio emitting X-ray binaries with relativistic jets, like
\object{SS~433}, \object{GRS~1915+105}, \object{GRO~J1655$-$40} or
\object{Cygnus~X-3}. The reader is referred to Mirabel \& Rodr\'{\i}guez
(\cite{mirabel99}) for a detailed review and to Castro-Tirado et~al.
(\cite{castro01}) and Durouchoux et~al. (\cite{durouchoux02}) for the most
recent studies of microquasars.

The $V=11.2$ star \object{LS~5039}, located at an estimated distance of
$\sim3$~kpc and close to the galactic plane ($l=16.88\degr$, $b=-1.29\degr$),
was initially proposed by Motch et~al. (\cite{motch97}) to be the optical
counterpart of the X-ray source \object{RX~J1826.2$-$1450}, a likely High Mass
X-ray Binary (HMXB). Using data from the NRAO VLA Sky Survey and follow-up VLA
observations, Mart\'{\i} et~al. (\cite{marti98}) discovered that the source
was also a non-thermal radio emitter with moderate variability. X-ray
observations of \object{RX~J1826.2$-$1450} carried out by Rib\'o et~al.
(\cite{ribo99}) showed that the X-ray spectrum was significantly hard up to
30~keV, with a strong Gaussian iron line at 6.6~keV, and neither pulsed nor
periodic emission was found on time scales of 0.02--2000~s and 2--200~days,
respectively. Paredes et~al. (\cite{paredes00}) discovered that the system
displays relativistic radio jets at milliarcsecond (mas) scales, revealing the
microquasar nature of \object{LS~5039}, and proposed an association with the
$\gamma$-ray source \object{3EG~J1824$-$1514}. While in the past the mass
donor had been classified as an O7V((f)) star, optical and near-IR
spectroscopic observations by Clark et~al. (\cite{clark01}) show that it is in
fact an O6.5V((f)) star. McSwain et~al. (\cite{mcswain01}) obtained the radial
velocity curve of the system, with their fitted parameters being a short
period of $P=4.117$~days, a high eccentricity of $e=0.41$, a radial velocity
of the system of $V_0=4.6$~km~s$^{-1}$ and a mass function of
$f(m)=0.00103~M_{\sun}$. Rib\'o et~al. (\cite{ribo02})
obtained optical and radio proper motions, 
and found that \object{LS~5039} is a runaway X-ray binary with a space
velocity $\sim150$~km~s$^{-1}$.
McSwain \& Gies (\cite{mcswain02}), based on wind accretion models, have
recently suggested a neutron star as the compact object in \object{LS~5039},
and proposed an inclination of $i\simeq30\degr$ for this binary system.

The persistent radio emission of \object{LS~5039} suggests that the jet is
always present. With the aim to confirm this hypothesis, and also to detect
the jet at larger angular scales than those imaged with the previous Very Long
Baseline Array (VLBA) measurements, we observed this microquasar with the
European VLBI Network (EVN) and the Multi-Element Radio-Linked Interferometer
Network (MERLIN). In this Letter we present the obtained results.

\begin{figure}[htpb]
\resizebox{\hsize}{!}{\includegraphics{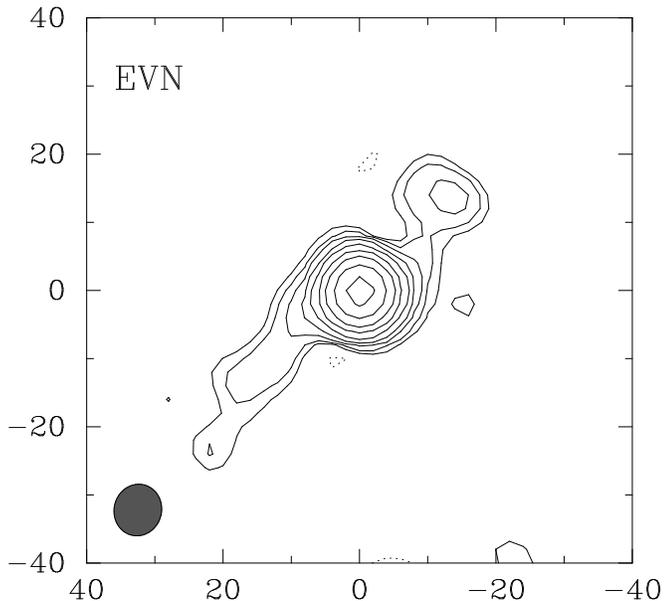}}
\caption{EVN self-calibrated image of \object{LS~5039} at 5~GHz obtained on March 1 2000. Axes units are in mas. The synthesized beam, plotted in the lower left corner, has a size of 7.60$\times$6.96~mas in P.A.\ of $-14\degr$. The first contour corresponds to 0.3~mJy~beam$^{-1}$, while consecutive ones scale with $3^{1/2}$.}
\label{fig:evn}
\end{figure}

\section{Observations and results} \label{sec:observations}

We observed \object{LS~5039} simultaneously with the EVN and MERLIN on March 1 
2000 (3:20--7:10~UT, MJD 51604.2) at 5~GHz.
The EVN observations were performed with 7 antennas, namely (antenna,
location, diameter)
Effelsberg, Germany, 100~m;
Jodrell Bank, UK, 25~m;
Cambridge, UK, 32~m;
Westerbork, The Netherlands, 14$\times$25~m;
Medicina, Italy, 32~m;
Noto, Italy, 32~m;
and
Toru\'n, Poland, 32~m.
Data were recorded in MkIV mode with 2-bit sampling at 256~Mbps at left hand
circular polarization, yielding a full bandwidth of 64~MHz.
The data were correlated at the Joint Institute for VLBI in Europe (JIVE).
Standard fringe fitting and imaging analysis were carried out using {\sc aips} and {\sc difmap}.
MERLIN recorded data with 2-bit sampling at dual polarization and a total
32~MHz bandwidth.
We analyzed the left hand circular polarization data excluding 1~MHz channel
at both edges of the band, yielding a final bandwidth of 14~MHz. 



\begin{figure}[htpb]
\resizebox{\hsize}{!}{\includegraphics{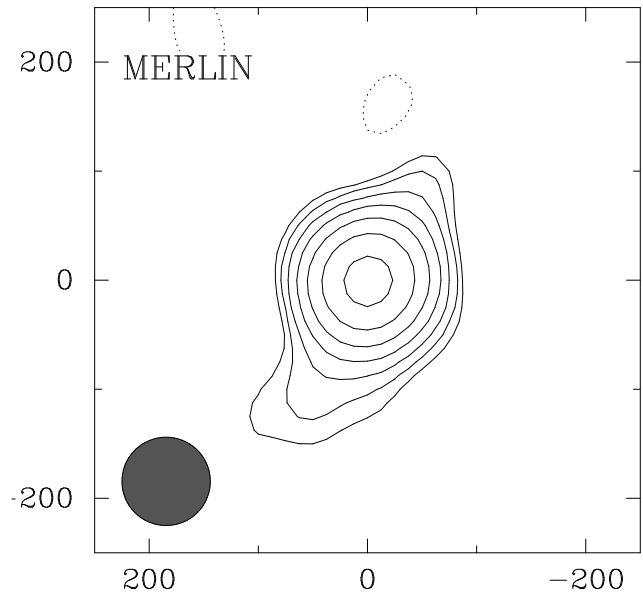}}
\caption{MERLIN self-calibrated image of \object{LS~5039} at 5~GHz obtained on March 1 2000. Axes units are in mas. The convolving circular beam, plotted in the lower left corner, has a diameter of 81~mas. The first contour corresponds to a flux density of 1~mJy~beam$^{-1}$, while consecutive ones scale with $3^{1/2}$.}
\label{fig:merlin}
\end{figure}

\begin{table}
\caption[]{Flux densities at 5~GHz, length and P.A.\ of the \object{LS~5039} structure features as detected by EVN and MERLIN.}
\begin{center}
\begin{tabular}{@{}l@{~~~}c@{~~~}c@{~~~}c@{~~~~~~~~}c@{~~~}c@{~~~}c@{}}
\hline \hline \noalign{\smallskip}
         & \multicolumn{3}{c}{EVN~~~~~~~} & \multicolumn{3}{c}{~~~MERLIN} \\
\noalign{\smallskip} \hline \noalign{\smallskip}
         & $S_{\rm \,5\,GHz}$ & Length & P.A.      & $S_{\rm \,5\,GHz}$ & Length & P.A. \\
         & [mJy]              & [mas]  & [$\degr$] & [mJy]              & [mas]  & [$\degr$] \\
\noalign{\smallskip} \hline \noalign{\smallskip}
Core     & 29.3               & ---    & ~~---     & 31.6               & ~---  & ~~--- \\
NW jet   & ~\,2.6             & 24     & $-$42     & ~\,4.0             & 128    & $-$29 \\
SE jet   & ~\,3.3             & 34     & ~140      & ~\,4.2             & 174    & ~150 \\
\noalign{\smallskip} \hline
\end{tabular}
\begin{list}{}{}
\footnotesize
\item[] Note: the errors in flux density, lenght and P.A.\ are
0.1~mJy~beam$^{-1}$, 2~mas and $4\degr$ for the EVN features and
0.4~mJy~beam$^{-1}$, 12~mas and $5\degr$ for the MERLIN ones.
\end{list}
\end{center}
\label{table:features}
\end{table}

Inspection of the shortest MERLIN baselines reveals a constant flux density
during the full run. That allowed us to image the source directly without
splitting the data in time blocks. We present the obtained EVN and MERLIN
images in Figs.~\ref{fig:evn} and \ref{fig:merlin}, respectively. The MERLIN
image is presented convolved with a circular beam, equivalent in solid angle
to the interferometric synthesized beam of 142$\times$46~mas at position angle
(P.A.) $-$47\degr. Our images clearly show that \object{LS~5039} has a bipolar
jet emanating from a central core. In both images the southeast (SE) jet is
brighter and larger than the northwest (NW) one, as can be seen in
Table~\ref{table:features}, where we have quoted the flux densities of the
core and the jets and the lengths and P.A.\ of the jets. Their total lengths
are $\sim60$~mas in the EVN image and $\sim300$~mas in the MERLIN one.

\section{Jet parameters and intrinsic sizes} \label{sec:jet}

These results confirm the existence of a two-sided radio jet in
\object{LS~5039} reported in previous VLBA 5~GHz observations by Paredes
et~al. (\cite{paredes00}). This source does not present any strong outburst
or, at least, none has been detected during the eleven month monitoring
carried out by the Green Bank Interferometer between 1998 September 16 and
1999 August 22 (Clark et~al. \cite{clark01}). On the other hand, inspection of
the RXTE All Sky Monitor data (Levine et~al. \cite{levine96}) reveals that the
X-ray flux was at the typical low level (Rib\'o et~al. \cite{ribo99}) at the
epoch of the EVN and MERLIN observations and several weeks before. All this
suggests that the jets are persistent, as the VLBI images obtained up to now
seem to indicate.

In the observations reported here, the jets extend further away than the
$\sim6$~mas of the VLBA observations at the same frequency. In all the images
the jets have similar position angles, $\sim125\degr$ in the VLBA image from
Paredes et~al. (\cite{paredes00}), $\sim140\degr$ in our EVN image, and
$\sim150\degr$ in our MERLIN image. These results suggest a bending of the
jets with increasing distance from the core and/or precession. 

The brightness and length asymmetry of the jet components may involve special
relativity effects (Mirabel \& Rodr\'{\i}guez \cite{mirabel99}). Hence,
assuming that this is the reason for the detected length asymmetry of the
jets, we can estimate some parameters by using the following equation:
\begin{equation}
\beta\cos\theta={{\mu_{\rm a}-\mu_{\rm r}}\over{\mu_{\rm a}+\mu_{\rm r}}}={{d_{\rm a}-d_{\rm r}}\over{d_{\rm a}+d_{\rm r}}}~,
\label{eqdist}
\end{equation}
where $\beta$ is the velocity of the jet flow in units of the speed of light,
$\theta$ is the angle between the direction of motion of the ejecta and the
line of sight and $\mu_{\rm a}$ and $\mu_{\rm r}$ are the proper motions of
the approaching (SE) and receding (NW) components, respectively. Although we
do not know the epoch of ejection of the terminal plasma, we can cancel the
time variable by using the relative distances to the core $d_{\rm a}$ and
$d_{\rm r}$, as expressed in Eq.~\ref{eqdist}. We will identify $d_{\rm a}$
and $d_{\rm r}$ with the lengths of the approaching and receding jet,
respectively. Using Eq.~\ref{eqdist} and the EVN values in
Table~\ref{table:features} we obtain $\beta\cos\theta=0.17\pm0.05$, and hence
$\beta>0.17\pm0.05$ and $\theta<80\degr\pm3\degr$. For the MERLIN values we
obtain $\beta>0.15\pm0.06$ and $\theta<81\degr\pm3\degr$. These values are
similar to those previously derived from the VLBA image, of
$\beta>0.15\pm0.04$ and $\theta<81\degr\pm2\degr$. We have not considered an
eventual larger size for the NW jet with a brightness below the image noise
level. McSwain \& Gies (\cite{mcswain02}) have recently proposed an
inclination of $i\simeq30\degr$ for \object{LS~5039}. If we assume that the
jet is perpendicular to the accretion disk, and that the disk lies in the
orbital plane of the binary system, then $\theta=i=30\degr$, and using the
values from the EVN image, we obtain $\beta=0.20\pm0.06$, which indicates a
mildly relativistic jet.

The total length of the EVN and MERLIN jets is $\sim60$ and $\sim300$~mas,
respectively. Considering that the source is located at 2.9~kpc (Rib\'o et~al.
\cite{ribo02}), these angular lengths translate into linear lengths in the
plane of the sky of $\sim175$ and $\sim870$~AU, respectively. Assuming that
$\theta=30\degr$ we obtain intrinsic total lengths of $\sim350$ and
$\sim$1740~AU, respectively, and lengths of the SE jet arm of $\sim200$ and
$\sim$1000~AU, respectively. Moreover, the jet width is smaller than one
synthesized beam even in the EVN image. This implies a jet half opening angle
$\leq6\degr$.

\section{A scenario based on the $\gamma$-ray/radio emission} \label{sec:scenario}

It is interesting to use the presence of 5~GHz synchrotron emitting particles
so far away in order to constrain the jet physical parameters. The starting
point of the following calculations will be the scenario for the proposed
EGRET emission of \object{LS~5039} presented in Paredes et~al.
(\cite{paredes00}). In this context, the high energy $\gamma$-ray photons
arise from inverse Compton (IC) scattering of UV photons from the luminous
companion star by the relativistic electrons that later, after having lost
part of their original energy due to IC losses, will account for the radio
emission in the jets. 

We will adopt here a very simple model of an expanding jet. Cylindrical
coordinates, $z$ and $r$ measured parallel and perpendicular to the jet axis,
are the best choice in our case. The jet is assumed to form at a distance
$z_0$ from the compact object and flows with a velocity $v=\beta\,c$. The
lateral expansion of the jet is parameterized as $r=r_0 (z/z_0)^{\epsilon}$,
where $r_0$ is the initial jet radius. A freely expanding conical jet would
correspond to $\epsilon=1$, while for slowed lateral expansion we would have
$\epsilon<1$ (see e.g. Hjellming \& Johnston, \cite{hjellming88}). The
expansion velocity perpendicular to the jet axis is thus
$v_{\perp}=dr/dt=\epsilon v r / z$. Concerning the magnetic field, we will
only consider the component perpendicular to the jet axis, because it has the
slowest decay if conservation of the magnetic flux is assumed. The magnetic
field along the jet will be thus parameterized as $B=B_0 (r/r_0)^{-1}=B_0
(z/z_0)^{-\epsilon}$.

A relativistic electron injected at the base of the jet will decrease its
energy mainly through adiabatic expansion, IC and synchrotron losses according
to:
\begin{equation}
\frac{d E}{d t} = - \frac{2}{3} \frac{v_{\perp} E}{r} 
-\alpha_{\rm IC} U_{\rm rad} E^2 -\alpha_{\rm S} B^2 E^2~.
\label{losses}
\end{equation}
The factor 2/3 in the adiabatic expansion term comes from the lateral
expansion of the jet. The constants $\alpha_{\rm IC}=3.97\times10^{-2}$ and
$\alpha_{\rm S}=2.37\times10^{-3}$ (c.g.s. units) are the coefficients of the
terms accounting for IC and synchrotron losses, respectively. The radiation
energy density $U_{\rm rad}$ is assumed to be dominated by stellar UV photons
from the O6.5V star, whose luminosity above 10~eV amounts to
$L_*\simeq5\times10^{38}$~erg~s$^{-1}$ and $U_{\rm rad}=L_*/[4\pi
c(a^2+z^2)]$. With the recently determined orbital parameters (McSwain et~al.
\cite{mcswain01}, McSwain \& Gies \cite{mcswain02}), the likely semimajor axis
of the orbit is $a=2.6\times10^{12}$~cm. Adopting these parameters, the
radiation energy density close to the compact object is so high that electrons
with energies higher than $10^{-3}$~erg will easily produce $\gamma$-ray
photons with energies of 100--1000~MeV, as detected by EGRET. Moreover,
integration of Eq.~\ref{losses} taking only into account the IC losses,
reveals that the energy of the electrons decays very fast to values of
$\sim5\times10^{-4}$~erg, regardless of their initial energy.

The IC contribution in Eq.~\ref{losses} should be comparable to adiabatic
losses at least at the base of the jet. Later on, the adiabatic expansion term
will soon become dominant due to its slower decay ($\propto z^{-1}$). 
Therefore, at the base of the jet we can request:
\begin{equation}
\alpha_{\rm IC} \frac{L_*}{4 \pi c (a^2+z_0^2)} E_0^2 \simeq \frac{2\epsilon}{3} \frac{v E_0}{z_0}~.
\end{equation}
We can use the values quoted above, together with $v=0.2c$ and an energy of
$E_0=5\times10^{-4}$~erg for the electrons injected at the base of the jet, to
solve for $z_0$ in this quadratic equation. The obtained results are 0.35,
0.85 and 1.30~AU for $\epsilon=1$, 1/2 and 1/3, respectively. The synchrotron
term can be shown to be negligible a posteriori and has not been considered
here.

Hereafter, we will proceed taking into account only the adiabatic expansion
term, because it is dominant compared to the IC one, and rewrite
Eq.~\ref{losses} as:
\begin{equation}
\frac{d E}{d t} \simeq -\frac{2\epsilon}{3} \frac{v E}{z}~.
\label{adia}
\end{equation}
This can be easily integrated to give $E \simeq E_0 (z/z_0)^{-2\epsilon/3}$.
As the energy decays, the synchrotron emission of relativistic electrons will
shift towards lower frequencies. In particular, the synchrotron frequency
expected at a distance $z$ is given by:
\begin{equation}
\nu_{\rm c} = 6.27 \times 10^{18} B E^2 \simeq 6.27 \times 10^{18} B_0 E_0^2 (z/z_0)^{-7\epsilon/3}~.
\end{equation}
The electrons still producing 6~cm ($\nu_{\rm c}=5\times 10^9$~Hz) synchrotron
emission at about $z=1000$~AU (the size of the MERLIN jets) are probably those
originally injected with energies of $E_0\simeq5\times10^{-4}$~erg. Then, the
magnetic field $B_0$ at the base of the jet can be roughly estimated as:
\begin{equation}
B_0 \simeq 8.0\times 10^{-10} E_0^{-2} \left[\frac{z}{z_0}\right]^{7\epsilon/3}~.
\end{equation} 
This provides values of $3.6\times10^5$, 12 and 0.6~G for $\epsilon=1$, $1/2$
and $1/3$, respectively. If magnetic flux is conserved, the corresponding
field at the end of the MERLIN jet would be $B_{1000\,{\rm AU}}=128$, $0.4$
and $0.06$~G. For comparison, the only independent estimates of the magnetic
field in the \object{LS~5039} jets come from simple equipartition arguments.
The results are merely indicative of the average magnetic field in the solid
angle of the sky covered by the jets. Using the total flux density and size of
the MERLIN jets given in Table~\ref{table:features} we derive a field value of
$>8\times 10^{-3}$~G, although this is only a lower limit because the jet
width is unresolved. The VLBA image by Paredes et~al. (\cite{paredes00}),
allowed them to estimate an approximate magnetic field of $\sim0.2$~G for
radio jets reaching up to a few mas. It is clear that a field of such
intensity is more consistent with the $\epsilon<1$ results than with the too
high magnetic fields implied by a simple conical jet ($\epsilon=1$).
Therefore, a slowly expanding jet, and hence well collimated, is favored to
power the \object{LS~5039} radio emission up to $\sim1000$~AU by the electrons
that previously contributed to the $\gamma$-ray emission, without {\it in
situ} acceleration being required. This is also in agreement with our upper
limit of $\leq6\degr$ for the jet half opening angle.

\section{Future prospects} \label{sec:future}

The high eccentricity of \object{LS~5039} provides both, a variable accretion
rate along the orbit implying a variable rate of electrons injected into the
jet, and a variable radiation energy density close to the compact object.
Hence, it is predicted a variability in the $\gamma$-ray luminosity correlated
with the orbital period, which may hopefully be detected by INTEGRAL and/or
GLAST. On the other hand, if the star exhibits intrinsic variations
in the UV photon flux, a variability in $\gamma$-rays would be seen correlated
to it.

Since \object{LS~5039} seems to have persistent radio jets, it would be
interesting to study the possibility of detecting Doppler-shifted lines. As
the jet flow has an estimated velocity of $\sim0.2c$, the amount of
Doppler-shift is expected to be comparable to that found in \object{SS~433}
(Kotani et~al. \cite{kotani96}; Marshall et~al. \cite{marshall02}). Hence,
spectroscopic X-ray observations with XMM or Chandra could be very useful to
constrain the physics of the relativistic flow and to study the matter content
within the jet. Finally, as pointed out by Levinson \& Waxman
(\cite{levinson01}) and by Distefano et~al. (\cite{distefano02}), if the jets
are hadronic we would expect the formation of TeV neutrinos, that could be
detected in the future with detectors of km$^2$-scale effective area.

Overall, these results confirm the existence of persistent relativistic radio
jets in \object{LS~5039}, indicating that this source is a very good target to
be studied with new instruments in order to gain knowledge on jet physics.

\begin{acknowledgements}

We acknowledge J.~Garc\'{\i}a-S\'anchez and A.~Lobanov for their useful
comments and suggestions after reading through a draft version of this
paper.
We acknowledge detailed and useful comments from L.~F. Rodr\'{\i}guez, the referee of this Letter.
We are very grateful to S.~T. Garrington, M.~A. Garrett, D.~C. Gabuzda, and C.~Reynolds for their valuable help in the data reduction process.
This paper is based on observations with the 100-m telescope of the MPIfR (Max-Planck-Institut f\"ur Radioastronomie) at Effelsberg.
We thank the staff of the JIVE correlator and of the observing telescopes.
The European VLBI Network is a joint facility of European, Chinese and other radio astronomy institutes funded by their national research councils.
MERLIN is operated as a National Facility by the University of Manchester at Jodrell Bank Observatory on behalf of the UK Particle Physics \& Astronomy Research Council.
The EVN observations were carried out thanks to the TMR Access to Large-scale Facilities programme under contract No. ERBFMGECT950012.
Part of the data reduction was done at JIVE with the support of the European Community - Access to Research Infrastructure action of the Improving Human Potential Programme, under contract No. HPRI-CT-1999-00045.
J.~M.~P., M.~R. and J.~M. acknowledge partial support by DGI of the Ministerio de Ciencia y Tecnolog\'{\i}a (Spain) under grant AYA2001-3092, as well as partial support by the European Regional Development Fund (ERDF/FEDER).
During this work, M.~R. has been supported by two fellowships from CIRIT (Generalitat de Catalunya, ref. 1998~BEAI~200293 and 1999~FI~00199).
J.~M. has been aided in this work by an Henri Chr\'etien International Research Grant administered by the AAS, and has been partially supported by the Junta de Andaluc\'{\i}a.

\end{acknowledgements}

\end{document}